\documentclass[aps,prl,twocolumn,superscriptaddress,floatfix,preprintnumbers]{revtex4-2}
\usepackage{amsmath}
\usepackage{bm}
\usepackage{graphicx}
\usepackage{amssymb}
\usepackage{mathtools}
\usepackage{multirow}
\usepackage{xcolor}
\usepackage[colorlinks,linkcolor=blue,citecolor=blue,urlcolor=blue]{hyperref}

\begin{document}
\preprint{RIKEN-iTHEMS-Report-26}

\title{Stochastic Quantization as Optimal Control}

\author{Lingxiao Wang}
\email{lingxiao.wang@riken.jp}
\affiliation{RIKEN Center for Interdisciplinary Theoretical and Mathematical Sciences (iTHEMS), Wako, Saitama 351-0198, Japan}
\affiliation{Institute for Physics of Intelligence, Graduate School of Science, The University of Tokyo, Bunkyo-ku, Tokyo 113-0033, Japan}

\date{\today}

\begin{abstract}
Stochastic quantization defines a Euclidean quantum field theory as the equilibrium of a fictitious-time Langevin dynamics, which reaches the Gibbs measure asymptotically. We show that this quantization can be formulated as a finite-time stochastic optimal control problem. A tractable reference process, naturally supplied by the free theory when available, provides an Ornstein--Uhlenbeck dynamics, while the full interaction enters as a reference-corrected terminal cost. The optimal control is a Doob-transform force that steers the path-reweighted terminal ensemble to the target at a prescribed time and for a given noise amplitude. A neural network learns the residual control, realizing this optimal stochastic quantization (OSQ). Because the path weights are exact, imperfect training increases the variance of estimators but does not introduce model bias. On multimodal potentials we recover all modes at finite time and find that the noise amplitude sets a practical diffusion-horizon window. In two-dimensional lattice scalar $\phi^4$ theory we recover observables from hybrid Monte Carlo simulations near the critical point. Quantization is thereby formulated as control rather than equilibration.
\end{abstract}

\maketitle

\paragraph{Introduction---}
Nonperturbative quantum field theory rests on the path integral, evaluated in practice from the Gibbs measure $p(\phi)\propto e^{-S[\phi]}$. Stochastic quantization (SQ), introduced by Parisi and Wu~\cite{Parisi:1980ys}, turns this measure into dynamics. The Euclidean field evolves along a fictitious time under a Langevin drift derived from the action. Wiener noise, together with its fluctuation--dissipation partner in the drift, carries the fluctuations that realize the Euclidean quantum measure rather than a classical saddle, and expectation values are recovered as equilibrium averages. The scheme requires no gauge fixing~\cite{Parisi:1980ys,Damgaard:1987rr,Namiki:1993fd} and extends to complex actions through complex Langevin dynamics~\cite{Parisi:1983mgm, Klauder:1983nn} (for modern developments, see Refs.~\cite{Berger:2019odf, Aarts:2026uiu}), yet it quantizes asymptotically. The finite Langevin time introduces a systematic bias, successive configurations remain correlated, and soft modes of the Langevin dynamics exhibit critical slowing down~\cite{Batrouni:1985jn}.

The same barriers hinder local Markov-chain Monte Carlo approaches, where critical slowing down~\cite{Wolff:1989wq} and topological freezing toward the continuum~\cite{Luscher2011} are infamous, and have motivated trivializing maps~\cite{Luscher:2009eq} and learned transports for simulating field theories on the lattice, e.g., normalizing flows and their continuous variants~\cite{Albergo:2019eim, Kanwar:2020xzo, Boyda:2020hsi, DelDebbio:2021qwf, Gerdes:2022eve, Caselle:2023mvh, Bauer:2024byr,Bonanno:2025pdp}, stochastic normalizing flows~\cite{Wu:2020snf, Caselle:2022acb, Bulgarelli:2024brv}, and diffusion models reinterpreted through stochastic quantization~\cite{Wang:2023exq, Fukushima:2024oij, Hirono:2024zyg, Aarts:2024rsl, Ranner:2024qtv, Zhu:2025pmw, Aarts:2025lpi, Aarts:2026zzr, Komijani:2026lan, Tomiya:2026tbc}; see Refs.~\cite{Cranmer:2023xbe, Aarts:2025gyp} for recent reviews. These advances establish finite-time learned transports as practical routes to Euclidean measures, and alternatively bring stochastic quantization into dialogue with renormalization group flows~\cite{Cotler:2022fze, Sheshmani:2025ylv, Cotler:2023lem} and optimal transport~\cite{Leonard:2014sc, Vargas:2023dds, Berner:2022oct}. That confluence raises a sharper question than sampling efficiency alone: can quantization itself be posed as a finite-horizon stochastic control problem, with Euclidean fields and quantum fluctuations as its defining ingredients?

\begin{figure}[t]
    \centering
    \includegraphics[width=\columnwidth]{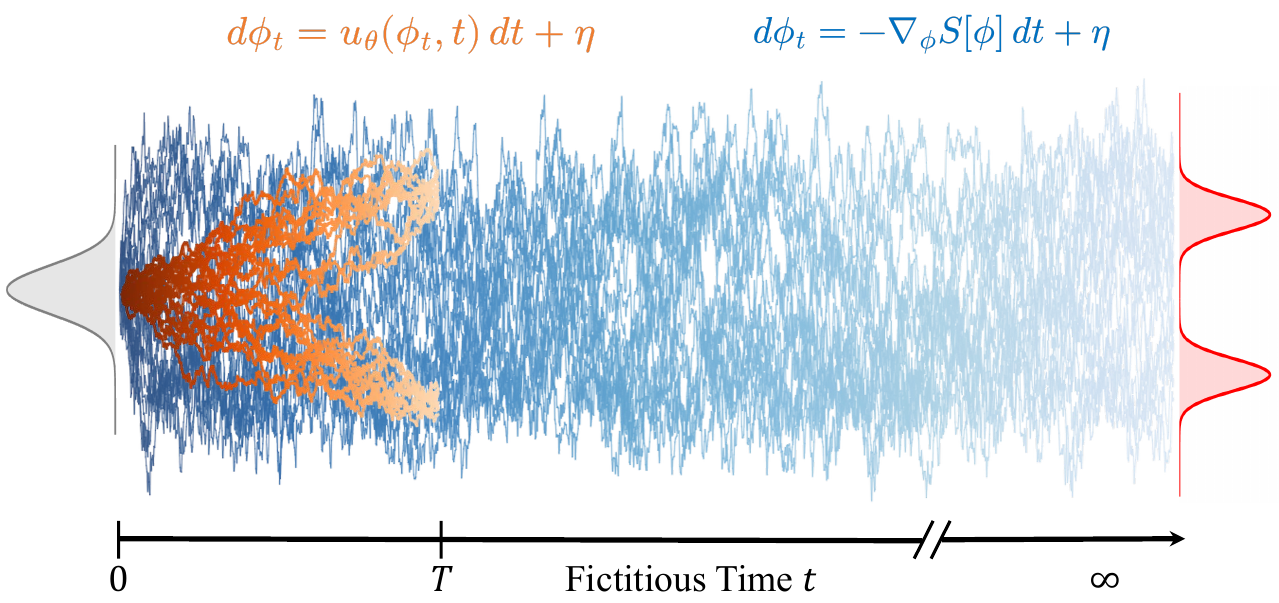}
    \caption{\label{fig:dds}
    Standard stochastic quantization (SQ) versus optimal stochastic quantization (OSQ).
    From the same prior (gray), uncontrolled Langevin dynamics (blue) reach the target
    $e^{-S}$ (red) only as $t\to\infty$; the controlled process with drift
    $u_\theta=f_0+v_\theta$ (orange) reaches a reweighted terminal ensemble for
    $e^{-S}$ at finite time $T$, each trajectory an independent weighted proposal.}
\end{figure}

In this Letter, we address quantization as a finite-horizon stochastic optimal-control problem, termed \emph{optimal stochastic quantization} (OSQ), whose solution targets the Gibbs measure at a prescribed fictitious time. Where a free theory exists, its Ornstein--Uhlenbeck drift supplies this reference and is retained as the physical backbone; more generally any analytically tractable reference process may play the same role. A neural network learns only the residual control that constructs the interacting theory to the target distribution at time $T$ (Fig.~\ref{fig:dds}). The Euclidean action enters the variational problem as a reference-corrected terminal cost, so equilibration is replaced by a finite-horizon terminal objective. The exactly reweighted ensemble follows $e^{-S}$ at finite time and noise amplitude, independently of training quality, and each trajectory delivers a statistically independent weighted proposal. We prove that the optimal residual force is a Doob transform~\cite{Doob:1984classical} of the free theory. Equivalently, the interacting terminal law can be represented as a Doob conditioning of free theory trajectories by their terminal Boltzmann weight, instead of being obtained through long-time equilibration.

What distinguishes OSQ is that the physical decomposition of the theory itself fixes the control ingredients: the free backbone transports soft and ultraviolet modes analytically, the network learns only a residual Doob force, and exactness is preserved by path weights. Generic controlled samplers instead learn the entire drift from a Brownian or Ornstein--Uhlenbeck reference unrelated to the theory~\cite{Zhang:2022pis, Vargas:2023dds, Berner:2022oct, Richter:2023isd}; closer methods add a learned drift to annealed Langevin dynamics with Jarzynski weights~\cite{2001Annealed, Albergo:2024nets}, or learn forward and backward drifts corrected by trajectory-level independence Metropolis--Hastings~\cite{Chen:2026bqw}. Physics-informed kernels~\cite{Ihssen:2025pik} share a similar philosophy by prescribing an analytic action path and solving the Wegner equation as a linear equation for the field kernel. Unlike deterministic continuous flows~\cite{Gerdes:2022eve}, the diffusion is retained; relative to Schr\"odinger bridges between prescribed marginals~\cite{Leonard:2014sc}, the reference and noise remain tied to the Euclidean action. The marginal Fisher information along fictitious time provides a partial sharpness diagnostic for the approach to $e^{-S}$.

\paragraph{Theory---}
Consider a field $\phi\in\mathbb{R}^n$ with Gibbs target
\begin{equation}
    p_{\mathrm{target}}(\phi)=\frac{1}{Z}\,e^{-S[\phi]},
    \qquad
    Z=\int\!d\phi\,e^{-S[\phi]}.
    \label{eq:gibbs}
\end{equation}
Parisi--Wu introduces fictitious time $t$ and the Langevin dynamics~\cite{Parisi:1980ys},
\begin{equation}
    d\phi_t=f(\phi_t)\,dt+g\,dw_t,\qquad
    f(\phi)\equiv-\frac{g^2}{2}\,\nabla_\phi S[\phi],
    \label{eq:langevin}
\end{equation}
whose Fokker--Planck equation admits $p_{\mathrm{target}}$ as the unique stationary law~\cite{Damgaard:1987rr, Namiki:1993fd, risken1996fokker}. The prefactor $g^2/2$ in the drift enforces fluctuation--dissipation balance, so that $p_{\mathrm{target}}$ is stationary for any $g>0$. Absent the Wiener term, the dynamics collapse toward classical minima of $S$, in contrast to deterministic microcanonical proposals~\cite{Callaway:1982eb, Polonyi:1983tm}. At fixed dimensionless $S\equiv S_E/\hbar$, changing $g$ in Eq.~\eqref{eq:langevin} is the fictitious-time reparametrization $\tau=g^{2}t/2$, so the equilibrium distribution is unchanged while the effective horizon becomes $g^{2}T/2$. The same holds for the controlled dynamics if the residual drift scales as $g^{2}$. Nonetheless $g$ still enters the fictitious-time path measure and therefore any finite-time importance weights built from it. Exact equality with the equilibrium distribution holds only as $t\to\infty$; when the Fokker--Planck operator is gapped, deviations exponentially decay~\cite{Berger:2019odf,Aarts:2024wxi}, so practical thermalization can be fast even though asymptotic exactness still requires infinite time.

To obtain finite-$T$ quantization while retaining an analytic backbone, we split $S=S_0+S_{\mathrm{int}}$ and take the free theory drift $f_0=-(g^2/2)\nabla_\phi S_0$. A residual control $v_\theta(\phi,t)=g^2\bar v_\theta(\phi,t)$ defines the controlled SDE
\begin{equation}
    d\phi_t=\bigl[f_0(\phi_t)+v_\theta(\phi_t,t)\bigr]\,dt+g\,dw_t,
    \label{eq:controlled_sde}
\end{equation}
with prior $q_0=\pi$ and marginals $q_t^\theta$. The task is to find the $v_\theta$ that transports $\pi$ to $p_{\mathrm{target}}$ at time $T$ with minimal kinetic cost relative to the free theory as a reference.

The instantaneous control-energy cost of the residual,
$\mathcal{J}(\theta,t)=\tfrac{1}{2g^{2}}\mathbb{E}_{q_t^\theta}[\|v_\theta\|^2]$,
integrates to the path objective
\begin{equation}
    \mathcal{L}_{\mathrm{ctrl}}(\theta)
    =\frac12\,\mathbb{E}_{\mathbb{P}_\theta}\!\left[
      \int_0^T g^{-2}\|v_\theta\|^2\,dt\right],
    \label{eq:fisher_int}
\end{equation}
which equals the path-measure KL divergence $\mathrm{KL}(\mathbb{P}_\theta\|\mathbb{P}_{\mathrm{ref}})$ by Girsanov's theorem~\cite{Oksendal:2003sde} (End Matter). Combining this path KL with the terminal action cost bounds $\mathrm{KL}(q_T^\theta\|p_{\mathrm{target}})$ from above (End Matter) yields the tractable control objective
\begin{equation}
    \mathcal{L}(\theta)
    =\mathbb{E}_{\mathbb{P}_\theta}\!\left[
      \frac12\int_0^T g^{-2}\|v_\theta\|^2\,dt
      +S_{\mathrm{eff}}[\phi_T]\right],
    \label{eq:final_loss}
\end{equation}
with the effective terminal potential
\begin{equation}
    S_{\mathrm{eff}}[\phi]\equiv S[\phi]+\log p_T^{\mathrm{ref}}(\phi),
    \label{eq:seff}
\end{equation}
and $p_T^{\mathrm{ref}}$ the closed-form Ornstein--Uhlenbeck marginal of the free reference. At the optimum,
$e^{-S_{\mathrm{eff}}}\,p_T^{\mathrm{ref}}=e^{-S}\propto p_{\mathrm{target}}$
is recovered at finite $T$.

Minimizing the loss~\eqref{eq:final_loss} is a standard stochastic optimal-control problem. Its Hamilton--Jacobi--Bellman equation implies the optimal residual
\begin{equation}
    \bar v^*(\phi,t)=\nabla_\phi\log h_t(\phi),
    \label{eq:optimal_v}
\end{equation}
or equivalently $v^*=g^2\nabla_\phi\log h_t$, where $h_t$ solves the free theory backward Kolmogorov equation with terminal condition $h_T=e^{-S_{\mathrm{eff}}}$. By Feynman--Kac~\cite{Karatzas:1991bm},
\begin{equation}
    h_t(\phi)
    =\mathbb{E}_{\mathbb{P}_{\mathrm{ref}}}\!\bigl[
      e^{-S_{\mathrm{eff}}[\phi_T]}\,\big|\,\phi_t=\phi\bigr],
    \label{eq:feynman_kac}
\end{equation}
so $h_t$ is the expected future Boltzmann weight along free theory trajectories. The optimal residual is therefore a Doob-transform force of the free theory (End Matter). At the optimum the ensemble at time $T$ coincides with the Gibbs measure up to a reweighting that depends only on the initial field; when that factor is constant over the prior support, the generated configurations already follow $e^{-S}$ at finite $T$ without reweighting.

\paragraph{Toy models---}
We first verify finite-$T$ quantization on two one-dimensional multimodal potentials. For the double well $S(\phi)=(\phi^2-1)^2$ we take $S_0\equiv0$, so $f_0\equiv0$ and the reference is pure diffusion $d\phi_t=g\,dw_t$. The barrier structure enters only through $v_\theta$ and $S_{\mathrm{eff}}$. For the sine-Gordon-type potential~\cite{Rajaraman:1982isy}
\begin{equation}
    S(\phi)=\tfrac12\phi^2-\cos(k\phi),
    \label{eq:sg_action}
\end{equation}
we split $S_0=\tfrac12\phi^2$ and $S_{\mathrm{int}}=-\cos(k\phi)$, yielding the Ornstein--Uhlenbeck backbone $f_0=-(g^2/2)\phi$ and controlled dynamics
\begin{equation}
    d\phi_t=\bigl[-\tfrac{g^2}{2}\phi_t+v_\theta(\phi_t,t)\bigr]\,dt+g\,dw_t.
    \label{eq:sg_sde}
\end{equation}
For both models the residual is a time-conditioned multilayer perceptron (MLP) constrained to be odd in the field, matching the even target potential; for the sine-Gordon-type model we use Fourier features with learnable frequencies. The residual is trained on the objective~\eqref{eq:final_loss} over the time $T=1$, discretized into $100$ uniform time steps, with the Adam optimizer. Exact target densities for comparison are obtained analytically.

\begin{figure}[t]
    \centering
    \includegraphics[width=\columnwidth]{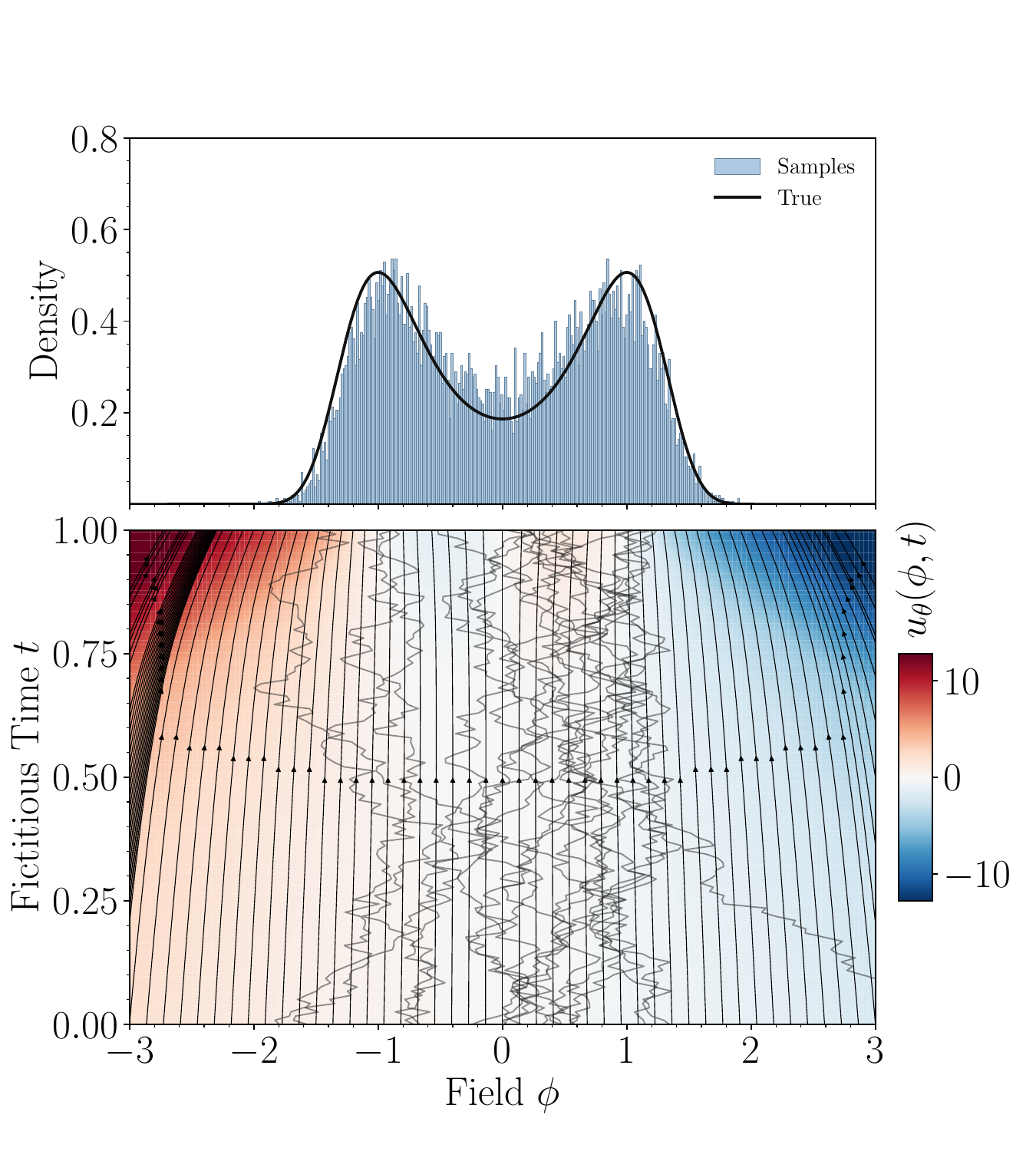}
    \caption{\label{fig:samples}
    Finite-$T$ distributions and learned drift for the double well $S=(\phi^2-1)^2$ at $g=1$.
    Top: $10^4$ OSQ samples (blue) versus the exact target (black); both modes
    at $\phi=\pm1$ are populated.
    Bottom: total drift $u_\theta=f_0+v_\theta$ (color) with sample trajectories (gray).
    The early landscape is barrier-free; the drift bifurcates into the two modes only as
    $t\to T$.}
\end{figure}

\begin{figure}[t]
    \centering
    \includegraphics[width=\columnwidth]{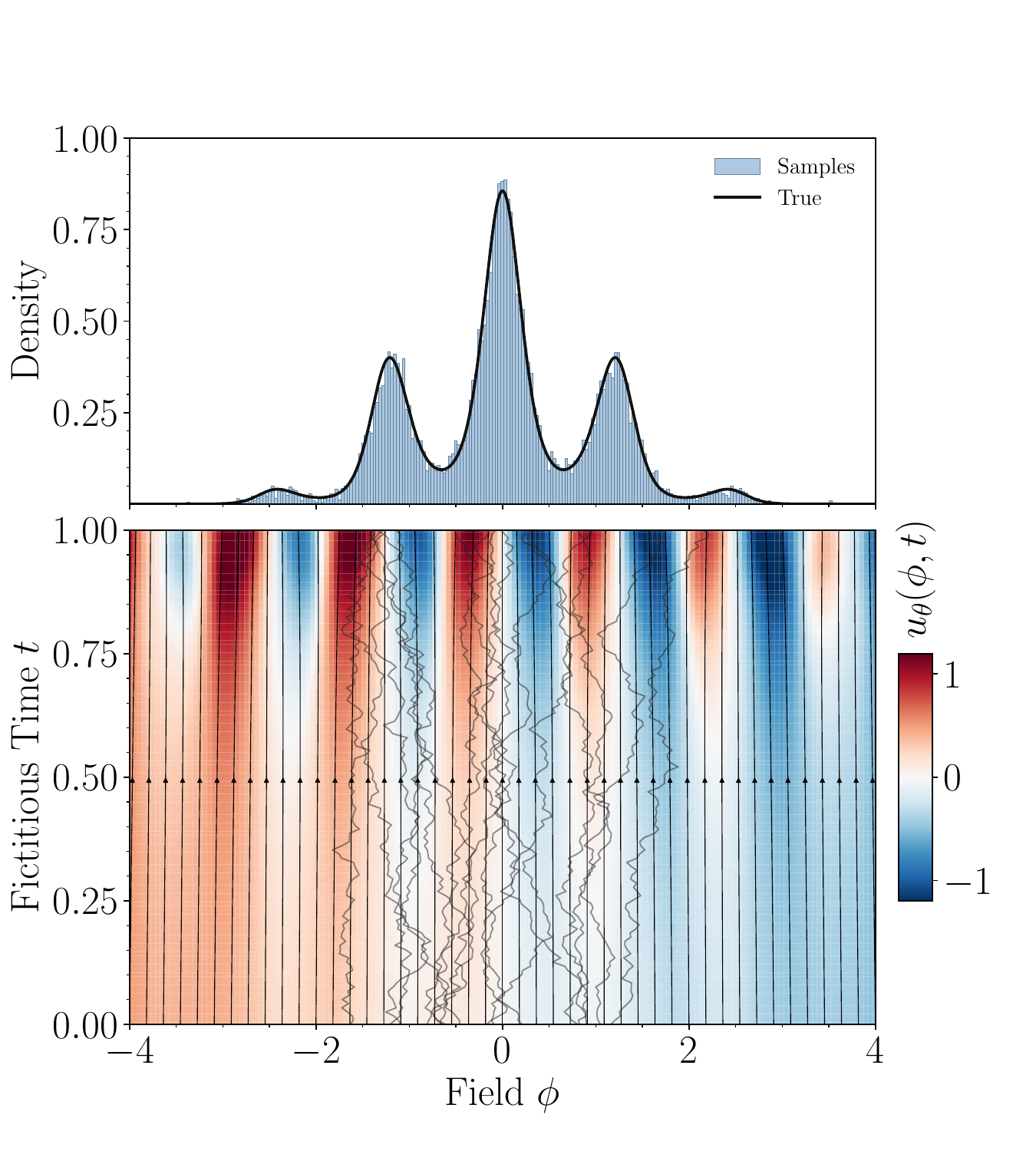}
    \caption{\label{fig:samples_sg}
    Same as Fig.~\ref{fig:samples}, for the sine-Gordon-type
    potential~\eqref{eq:sg_action} with $k=5$ at $g=0.5$.
    Top: all symmetric peaks and the central maximum are reproduced.
    Bottom: the total drift funnels mass into the local minima.}
\end{figure}

\begin{figure}[hbpt!]
    \centering
    \includegraphics[width=\columnwidth]{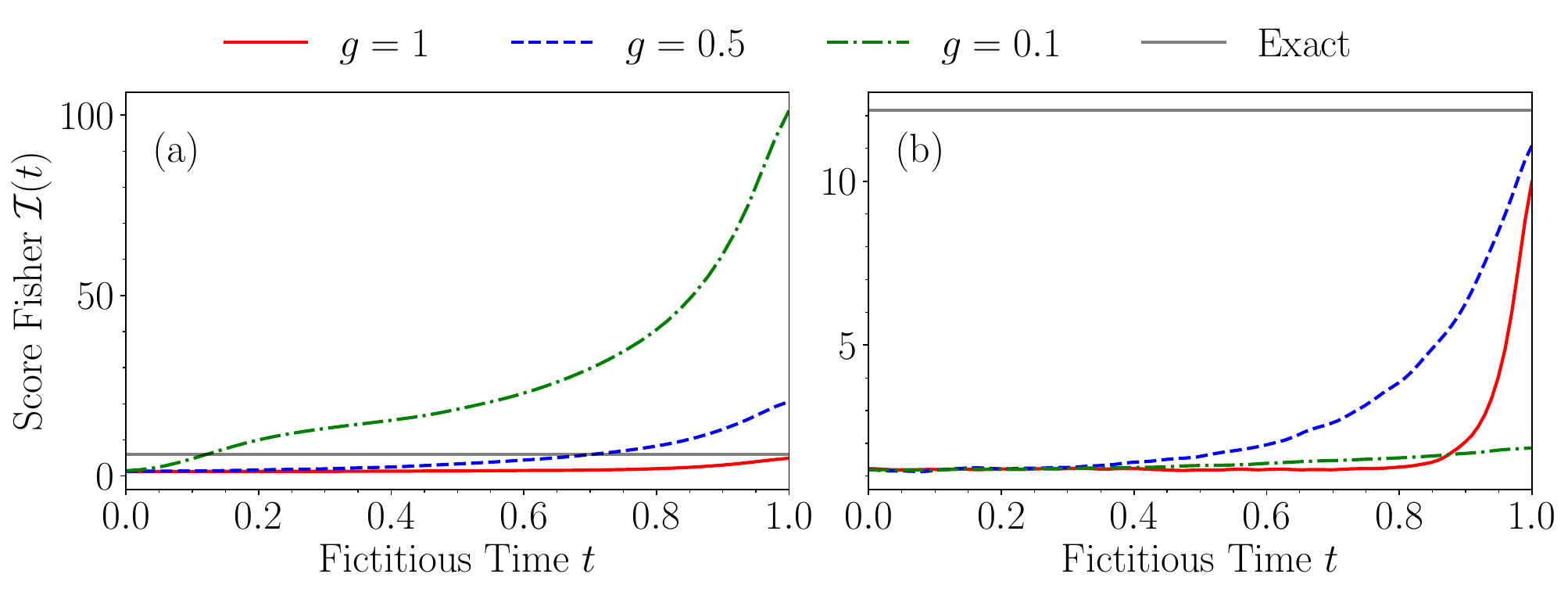}
    \caption{\label{fig:cfunc}
    Marginal Fisher information $\mathcal{I}(t)$ along fictitious time for OSQ at
    $g=1$, $0.5$, and $0.1$, for (a)~the double well and (b)~the sine-Gordon-type
    model. Gray lines mark the $g$-independent target
    $\mathbb{E}_{p_{\mathrm{target}}}[\|\nabla S\|^2]$ ($5.99$ and $12.16$).
    $\mathcal{I}(t)$ grows monotonically for all values of $g$ studied, but reaches the target only
    inside the practical noise window (see text).}
\end{figure}

Figures~\ref{fig:samples} and \ref{fig:samples_sg} (top) show that the finite-$T$ marginals reproduce both target distributions. To diagnose how the noise shapes the approach of the finite-time ensemble to target, we monitor the marginal Fisher information
$\mathcal{I}(t)=\mathbb{E}_{q_t}[\|\nabla\log q_t\|^2]$,
the mean squared score of $q_t$. As probability mass organizes onto the modes of $S$, $\mathcal{I}(t)$ increases; we estimate it by Gaussian kernel density estimation with analytic score~\cite{Silverman:1986kde}, refined by Richardson extrapolation~\cite{Richardson:1927}. For each amplitude studied, $\mathcal{I}(t)$ grows monotonically toward the terminal time (Fig.~\ref{fig:cfunc}), the reverse of the de~Bruijn decay under pure diffusion~\cite{Cover:2006elements}. For the target itself,
$\mathcal{I}=\mathbb{E}_{p_{\mathrm{target}}}[\|\nabla S\|^2]$
($5.99$ and $12.16$ for the two models); whether $\mathcal{I}(T)$ attains this value exposes a two-sided role of the noise amplitude.

\begin{table*}[hbpt!]
    \caption{\label{tab:phi4_comparison}%
        OSQ versus HMC observables in two-dimensional scalar $\phi^4$ theory at
        $\lambda=0.022$.
        OSQ uses $N\simeq10^4$ independent trajectories at $g=\sqrt{2}$,
        reweighted by the exact path weights of Eq.~\eqref{eq:girsanov};
        HMC values are from Ref.~\cite{Tan:2026ibs}.
        Parentheses give uncertainties on the last digit(s);
        the last column is $\mathrm{ESS}/N$ for the reweighted OSQ ensemble.
    }
    \footnotesize
    \begin{ruledtabular}
    \begin{tabular}{cl cc cc cc cc c}
            & & \multicolumn{2}{c}{$\langle|M|\rangle$}
            & \multicolumn{2}{c}{$\chi$}
            & \multicolumn{2}{c}{$\langle S\rangle/V$}
            & \multicolumn{2}{c}{$U_L$} & \\
            $L$ & $\kappa$ & HMC & OSQ & HMC & OSQ & HMC & OSQ & HMC & OSQ & $\mathrm{ESS}/N$ \\
            \colrule
            \multirow{3}{*}{$16$}
            & $0.26$   & $0.254(2)$  & $0.250(2)$ & $8.4(1)$   & $8.5(2)$ & $0.4511(4)$ & $0.4494(5)$ & $0.132(9)$  & $0.116(13)$ & $0.64$ \\
            & $0.2705$ & $0.807(3)$  & $0.793(7)$ & $23.6(3)$  & $24.6(9)$ & $0.3883(5)$ & $0.3895(10)$ & $0.548(2)$  & $0.541(4)$ & $0.27$ \\
            & $0.28$   & $1.455(1)$  & $1.446(8)$ & $4.23(8)$  & $4.3(4)$ & $0.2388(5)$ & $0.238(3)$ & $0.6569(2)$ & $0.6563(9)$ & $0.024$ \\
            \colrule
            \multirow{3}{*}{$32$}
            & $0.26$   & $0.1250(9)$ & $0.128(3)$ & $9.0(1)$   & $9.3(4)$ & $0.4517(2)$ & $0.4514(5)$ & $0.02(1)$   & $0.04(3)$ & $0.20$ \\
            & $0.2705$ & $0.718(2)$  & $0.725(19)$ & $63.9(10)$ & $67(5)$ & $0.3931(3)$ & $0.3914(17)$ & $0.568(2)$  & $0.568(10)$ & $0.014$ \\
            & $0.28$   & $1.4561(6)$ & $1.454(5)$ & $4.11(6)$ & $3.8(5)$ & $0.2383(2)$ & $0.238(3)$ & $0.66417(4)$ & $0.6643(3)$ & $0.006$ \\
            \colrule
            \multirow{3}{*}{$64$}
            & $0.26$   & $0.0632(5)$ & $0.066(4)$ & $9.0(2)$   & $9.4(1.1)$ & $0.4511(1)$ & $0.4525(9)$ & $0.05(1)$   & $0.07(11)$ & $0.013$ \\
            & $0.2705$ & $0.600(2)$  & $0.52(4)$ & $213(3)$   & $170(37)$ & $0.3964(1)$ & $0.3988(26)$ & $0.554(2)$  & $0.53(4)$ & $0.002$ \\
            & $0.28$   & $1.4556(3)$ & $1.463(12)$ & $4.07(6)$  & $5.5(2.0)$ & $0.2382(1)$ & $0.237(3)$ & $0.666043(9)$ & $0.6658(3)$ & $0.0007$ \\
    \end{tabular}
    \end{ruledtabular}
\end{table*}

The noise must be strong enough to populate separated modes. At $g=0.1$ secondary minima of the sine-Gordon-type potential remain unresolved, while the double-well peaks at $\phi=\pm1$ become narrower than the target, consistent with the near-deterministic limit in which trajectories collapse toward classical minima~\cite{Damgaard:1987rr}. Restoring the peak structure of the target nonetheless becomes harder as $g$ grows, because the residual force needed to hold a peak of width $\delta$ against diffusion scales as $g^{2}/\delta^{2}$ near $t=T$. At $g=1$ the sine-Gordon-type marginal is the target convolved with a Gaussian of width $\sigma\approx0.07$, saturating $\mathcal{I}(T)\simeq10.7$; at $g=0.5$, $\mathcal{I}(T)$ lies closest to target. Thus $g$ sets a practical diffusion-horizon window at fixed $T$. The learned drift channels configurations through intermediate times where barriers are softened before the mode structure hardens at $t=T$ (Figs.~\ref{fig:samples} and \ref{fig:samples_sg}, bottom).

\paragraph{Lattice $\phi^4$ theory---}
We next quantize two-dimensional lattice scalar $\phi^4$ theory~\cite{Tan:2026ibs} at finite fictitious time. The model has $\mathbb{Z}_2$ symmetry $\phi\to-\phi$ and a second-order transition in the Ising universality class. In the standard dimensionless parametrization on an $L\times L$ torus (End Matter), the lattice action arrives directly in the free-plus-interaction form $S=S_0+S_{\mathrm{int}}$ of the OSQ decomposition,
\begin{align}
    S_0[\phi]&=\sum_x\Big[
        -2\kappa\sum_{\mu=1}^{2}\phi(x)\,\phi(x+\hat\mu)
        +(1-2\lambda)\phi^2(x)\Big],
    \nonumber\\
    S_{\mathrm{int}}[\phi]&=\sum_x\lambda\,\phi^4(x),
    \label{eq:phi4_decomp}
\end{align}
with hopping parameter $\kappa$ and quartic coupling $\lambda$. The free theory backbone $f_0=-(g^2/2)\nabla_\phi S_0$ is an Ornstein--Uhlenbeck process diagonal in momentum space, with mode stiffness $K(p)=(1-2\lambda)-2\kappa\sum_\mu\cos p_\mu$. The quartic force $-4\lambda\phi^3$ enters only through $S_{\mathrm{eff}}$ and the learned residual $v_\theta$. For fixed $\lambda$, the symmetric and broken phases are separated by $\kappa_c(\lambda)$. Classically $\kappa_c^{\mathrm{cl}}=(1-2\lambda)/(2d)$.

For $\kappa>\kappa_c^{\mathrm{cl}}$ the infrared stiffness $K(0)$ turns negative, since the split~\eqref{eq:phi4_decomp} places the inverted quadratic piece in $S_0$. The free Ornstein--Uhlenbeck process then has no stationary measure; its soft-mode instability seeds spontaneous symmetry breaking. OSQ can use the finite-$T$ Gaussian marginal of this unstable reference, but a stable backbone conditions the control problem better. For example, a mass shift restores $K(p)>0$ with the reference wells kept soft. In the broken phase we expand about the mean-field vacua $\pm\mu_{\mathrm{MF}}$, $\mu_{\mathrm{MF}}^2=-K(0)/(2\lambda)$, as a $\mathbb{Z}_2$-symmetric two-center Gaussian mixture, the lattice analogue of the separated modes in the toy models, whose exact marginal enters $S_{\mathrm{eff}}$ and whose centers are drawn by the prior at equal weight. Near criticality, where the magnetization is broad and only weakly bimodal, the wells are drawn inward and softened with volume following two-dimensional Ising finite-size scaling (End Matter). The residual $v_\theta$ absorbs both $S_{\mathrm{int}}$ and the mismatch between $S_0$ and this stabilized reference, while path reweighting remains exact.

We fix $\lambda=0.022$ and $g=\sqrt{2}$ and test the observables with the hybrid Monte Carlo approach~\cite{Tan:2026ibs,Duane:1987de} at $L=16,32,64$ and $\kappa=0.26,0.2705,0.28$, with parameters across the critical point (Table~\ref{tab:phi4_comparison}). The network architecture, discretization, and training budget are kept fixed across all $(L,\kappa)$; details are given in the End Matter. The purpose of this comparison is to test finite-time quantization rather than algorithmic speed. The ratio $\mathrm{ESS}/N$ quantifies the degradation of overlap with volume in the ordered phase, the generic bottleneck of global transport methods.

Standard observables are the absolute magnetization
$\langle|M|\rangle$ with $M=V^{-1}\sum_x\phi(x)$ and $V=L^2$,
the susceptibility $\chi=V(\langle M^2\rangle-\langle|M|\rangle^2)$,
the action density $\langle S\rangle/V$,
and the Binder cumulant $U_L=1-\langle M^4\rangle/(3\langle M^2\rangle^2)$~\cite{Binder:1981zz}. As a finite-volume proxy for the order parameter, $\langle|M|\rangle>0$ and $U_L\to2/3$ as $L\to\infty$ in the broken phase, whereas both vanish in the symmetric phase.

Each trajectory carries an exact importance weight
$\log w_i=-S_{\mathrm{eff}}[\phi_T^{(i)}]-\log\bigl(d\mathbb{P}_\theta/d\mathbb{P}_{\mathrm{ref}}\bigr)$,
built from the reference-corrected terminal action and the path correction of
Eq.~\eqref{eq:girsanov}. Reweighting with these weights yields consistent estimates
of observables in the large-sample limit whenever the controlled dynamics explores the full configuration
space; what is specific to OSQ is that the weight is known in closed form for
every trajectory and, at the optimum, depends only on the prior draw (End Matter).
Incomplete training then reduces the effective sample size rather than introducing
model bias.

At $L=16$, and in particular near $\kappa=0.2705$, reweighted OSQ agrees with HMC for the tabulated observables within combined errors, including the Binder cumulant, with $\mathrm{ESS}/N=0.27$ near criticality and $0.64$ in the symmetric phase. The agreement continues through $L=32$. At $L=64$, where $\mathrm{ESS}/N$ is smallest, the uncertainties enlarge and the comparison loses precision, especially near $\kappa=0.2705$. Without weights the susceptibility can be misestimated, while path reweighting restores HMC agreement at reduced ESS (e.g.\ $0.024$ at $L=16$, $\kappa=0.28$). With fixed training budget, $\mathrm{ESS}/N$ decreases with volume and toward the broken phase as the residual Doob force grows. The open frontier is therefore finite-$T$ control of that interacting residual.

\paragraph{Discussion---}
In OSQ the free theory stochastic process supplies the physical backbone, the Euclidean action enters as the terminal cost, and the learned residual provides the Doob force that completes the transport at finite fictitious time $T$. Exactness follows from the path weights at finite time rather than from long-time equilibration, and independent trajectories replace thermalization along a single Markov chain. Bare $S_0$ has no interaction barrier; the broken-phase mixture is bimodal only through a prior sector choice, with each conditional reference an Ornstein--Uhlenbeck process that need not tunnel, while residual control and path weights restore the interacting target. On the multimodal potentials, Wiener noise populates separated modes at intermediate times before the target barriers harden, which requires an amplitude window set by mode mixing versus restoration of the target peaks.

The construction places OSQ on the stochastic side of the duality between quantization and generative models. At a formal level, the deterministic/stochastic distinction parallels that between Bohmian and Nelson-type trajectory formulations~\cite{Bohm:1951xw, Fenyes:1952zph, Nelson:1966sm}, here drawn in fictitious rather than physical time. After training, independent draws of the initial field and of the Wiener noise produce statistically independent trajectories that can be integrated in parallel; for fixed lattice volume and fictitious-time discretization the cost per trajectory is fixed, so the cost of learning the residual is shared across large ensembles. The open frontier is the growth of the residual Doob dressing with volume in ordered phases (Table~\ref{tab:phi4_comparison}), which calls for volume-adapted references and training budgets. Important future directions include gauge theories with topological sectors, where freezing limits local updates~\cite{Luscher2011}; complex actions, where the dynamics connects to complex Langevin methods~\cite{Berger:2019odf, Aarts:2026uiu} and to physics-informed kernels for sign problems~\cite{Ihssen:2026sik}; and scaling toward criticality~\cite{Tan:2026ibs}. In the scalar theory the free backbone relaxes Fourier modes at rates set by the quadratic kernel $K(p)$, so fictitious time also admits a scale-dependent reading. This invites a comparison with diffusion-based and renormalization-group constructions~\cite{Cotler:2023lem, Sheshmani:2025ylv, Masuki:2025rgdm, Cotler:2022fze, Carosso:2019qpb, Oh:2012bx}, although a precise correspondence is left for future work.

\paragraph{Code availability---}
An open-source code with scripts reproducing all results of this Letter, is publicly available at \url{https://github.com/DM-QFT/OSQ-Scalar}.

\paragraph{Use of AI tools---}
The initial ideas of this work took shape in early discussions with Gemini 3 Pro (Google). Claude Opus 4.8 and Claude Fable 5 (Anthropic) assisted substantially with numerical cross-checks and with polishing the manuscript. All derivations, results, and statements were verified by the author, who takes full responsibility for the content.

\begin{acknowledgments}
We thank Gert Aarts and Koji Hashimoto for insightful suggestions, Yang-yang Tan for numerical suggestions, Xu Feng, S. Prem Kumar, Jan Pawlowski, Lei Wang, Hong-An Zeng and all DM-QFT collaboration members for helpful discussions.
We thank the DEEP-IN working group at RIKEN-iTHEMS for support in the preparation of this paper.
L.W. is supported by the RIKEN-TRIP initiative (RIKEN-Quantum), JSPS KAKENHI Grant No. 25H01560, and JST-BOOST Grant No. JPMJBY24H9.
\end{acknowledgments}

\bibliographystyle{apsrev4-2}
\bibliography{ref}

\clearpage
\onecolumngrid
\begin{center}
  \textbf{\large End Matter}
\end{center}
\vspace{0.5em}
\twocolumngrid

\appendix

\section{Path-measure objective}
\label{app:fisher}

Let $\mathbb{P}_\theta$ be the path measure of the controlled
SDE~\eqref{eq:controlled_sde} and $\mathbb{P}_{\mathrm{ref}}$ that of
the free theory reference $d\phi_t=f_0(\phi_t)\,dt+g\,dw_t$ started
from $p_0=\pi$, with marginals $\{p_t\}_{t\in[0,T]}$; we abbreviate
$p_T\equiv p_T^{\mathrm{ref}}$. Both processes share the diffusion $g$
and the prior $\pi$, so the two path measures are mutually absolutely
continuous, and Girsanov's theorem~\cite{Oksendal:2003sde}, expressed
along controlled paths via the $\mathbb{P}_\theta$-Brownian motion
$d\tilde w_t = dw_t^{\mathrm{ref}} - g^{-1}v_\theta\,dt$, gives
\begin{equation}
    \log\frac{d\mathbb{P}_\theta}{d\mathbb{P}_{\mathrm{ref}}}
    = \int_0^T g^{-1}v_\theta\cdot d\tilde w_t
      + \frac{1}{2}\int_0^T g^{-2}\|v_\theta\|^2\,dt,
    \label{eq:girsanov}
\end{equation}
whose It\^o term is a $\mathbb{P}_\theta$-martingale and vanishes in
expectation. Hence
\begin{equation}
    \mathrm{KL}(\mathbb{P}_\theta\|\mathbb{P}_{\mathrm{ref}})
    = \frac{1}{2}\,\mathbb{E}_{\mathbb{P}_\theta}\!\left[
        \int_0^T g^{-2}\|v_\theta\|^2\,dt\right]
    = \mathcal{L}_{\mathrm{ctrl}}(\theta).
    \label{eq:fisher_eq_kl}
\end{equation}

\section{Terminal cost and finite-time exactness}
\label{app:loss}

The true objective is
$\mathcal{L}_{\mathrm{true}}=\mathrm{KL}(q_T^\theta\|p_{\mathrm{target}})
=\mathbb{E}_{q_T^\theta}[\log q_T^\theta]+\mathbb{E}_{q_T^\theta}[S]+\log Z$.
The entropy term has no closed form, but the chain rule of the path KL,
decomposed from the terminal end, gives
$\mathrm{KL}(\mathbb{P}_\theta\|\mathbb{P}_{\mathrm{ref}})
=\mathrm{KL}(q_T^\theta\|p_T)$ plus the (non-negative) expected KL of
the path measures conditioned on $\phi_T$, hence
$\mathrm{KL}(q_T^\theta\|p_T)\le
\mathrm{KL}(\mathbb{P}_\theta\|\mathbb{P}_{\mathrm{ref}})$, i.e.
$\mathbb{E}_{q_T^\theta}[\log q_T^\theta]\le
\mathbb{E}_{q_T^\theta}[\log p_T]+\mathcal{L}_{\mathrm{ctrl}}$.
Substituting into $\mathcal{L}_{\mathrm{true}}$ yields, up to constants,
\begin{equation}
    \mathcal{L}_{\mathrm{true}}(\theta)
    \le \mathcal{L}_{\mathrm{ctrl}}(\theta)
    + \mathbb{E}_{q_T^\theta}\!\bigl[S_{\mathrm{eff}}[\phi_T]\bigr],
    \label{eq:upper_bound}
\end{equation}
with $S_{\mathrm{eff}}=S+\log p_T$ as in Eq.~\eqref{eq:seff}; this is the
objective~\eqref{eq:final_loss}. For the Ornstein--Uhlenbeck reference
$p_T$ is an explicit Gaussian (in the broken phase, a two-component
mixture), so $S_{\mathrm{eff}}$ costs nothing extra to evaluate.
As the reference relaxes, $\log p_T\to -S_0$ and
$S_{\mathrm{eff}}\to S_{\mathrm{int}}$ up to constants, while
exactness at finite $T$ relies only on the exact marginal $p_T$.

\emph{Optimal control and Doob transform.}---For the value function
$\mathcal{V}(\phi,t)=\min_{v}\mathbb{E}[\frac12\int_t^T g^{-2}\|v\|^2 ds
+S_{\mathrm{eff}}[\phi_T]\,|\,\phi_t=\phi]$, the Hamilton--Jacobi--Bellman
equation is minimized pointwise by $v^*=-g^2\nabla_\phi \mathcal{V}$, and the
Cole--Hopf substitution $\mathcal{V}=-\log h_t$ linearizes it into the backward
Kolmogorov equation
\begin{equation}
    \partial_t h_t + \mathcal{L}_{f_0} h_t = 0,
    \qquad h_T = e^{-S_{\mathrm{eff}}},
    \label{eq:backward_pde_app}
\end{equation}
with $\mathcal{L}_{f_0}=f_0\cdot\nabla_\phi+\frac{g^2}{2}\Delta_\phi$,
so $v^*=g^2\nabla_\phi\log h_t$ with the Feynman--Kac
representation~\cite{Karatzas:1991bm}
$h_t(\phi)=\mathbb{E}_{\mathbb{P}_{\mathrm{ref}}}
[e^{-S_{\mathrm{eff}}[\phi_T]}|\phi_t=\phi]$. The optimal residual is
thus the Doob-transform force~\cite{Doob:1984classical} of the free theory.

\emph{Terminal law.}---Under the optimal drift
$u^*=f_0+g^2\nabla_\phi\log h_t$, the ansatz $q_t^*=h_t\,r_t$ in the
Fokker--Planck equation cancels the $\partial_t h_t$ terms by
Eq.~\eqref{eq:backward_pde_app} and leaves for $r_t$ exactly the
Fokker--Planck equation of the reference process, started from
$r_0=\pi/h_0$.
Moreover, It\^o's formula along the optimal trajectories gives
$\log(d\mathbb{P}_{\theta^*}/d\mathbb{P}_{\mathrm{ref}})
=\log h_T(\phi_T)-\log h_0(\phi_0)$, so the main-text weight
$\log w=-S_{\mathrm{eff}}-\log(d\mathbb{P}_{\theta}/d\mathbb{P}_{\mathrm{ref}})$
collapses at the optimum to $\log w=\log h_0(\phi_0)$. It depends on
the prior draw alone, and the optimal effective sample size is
$[\mathbb{E}_\pi h_0]^2/\mathbb{E}_\pi[h_0^2]$.
Whenever $h_0$ is constant on the support of $\pi$ (for a concentrated
prior, or once the reference mixes over the time), $r_t\propto p_t$
and
\begin{equation}
    q_T^* \propto e^{-S_{\mathrm{eff}}}\,p_T = e^{-S},
    \label{eq:terminal_marginal}
\end{equation}
so the unweighted optimal law equals $p_{\mathrm{target}}$ at finite
$T$ for any $g$; for a general prior the weights $w=h_0(\phi_0)$
restore Eq.~\eqref{eq:terminal_marginal} identically.
If the uncorrected cost $S$ is used instead ($h_T=e^{-S}$), the same
steps give $q_T^*\propto e^{-S}\,p_T$, exact only when $p_T$ is constant
over the relevant support; the reference correction removes this reliance.

\section{Lattice $\phi^4$ setup and training details}
\label{app:phi4}

The action~\eqref{eq:phi4_decomp} is the standard dimensionless
lattice form of the continuum theory
$S_E^{\rm cont}=\int d^2x\,[\tfrac12(\partial_\mu\phi_0)^2
+\tfrac12 m_0^2\phi_0^2+\tfrac{\lambda_0}{4!}\phi_0^4]$
on an $L\times L$ torus of spacing $a$, with
$a^{(d/2-1)}\phi_0=(2\kappa)^{1/2}\phi$,
$\lambda=a^{4-d}\kappa^2\lambda_0/6$, and
$(am_0)^2=(1-2\lambda)/\kappa-2d$ at
$d=2$~\cite{Gattringer:2010zz}. The backbone
$f_0=-(g^2/2)\nabla_\phi S_0$ is the hopping drift
$f_0(x)=g^2[\kappa\sum_{\mu}(\phi(x+\hat\mu)+\phi(x-\hat\mu))
-(1-2\lambda)\phi(x)]$.

\emph{Stabilized reference.}---For $\kappa>\kappa_c^{\mathrm{cl}}$ the
mass shift restoring $K(p)>0$ is $\max[-2K(0),0]+\varepsilon$, where
the residual well curvature $\varepsilon$ defines a reference
susceptibility $\chi_{\mathrm{ref}}=1/\varepsilon$. This stabilized
process replaces the bare $S_0$ backbone in the controlled SDE.
Deep in the broken phase the mixture wells sit near
$\pm\mu_{\mathrm{MF}}$, with stiff curvature; near
criticality they are drawn inward and softened with volume following
two-dimensional Ising finite-size scaling (separation
$\propto L^{-\beta/\nu}$, $\chi_{\mathrm{ref}}\propto L^{\gamma/\nu}$,
with $\beta/\nu=1/8$, $\gamma/\nu=7/4$), so that the reference tracks the
physical magnetization and susceptibility; in the symmetric phase a
single Gaussian at the origin suffices; the full
$(L,\kappa,\lambda)$ map is given in the released code.

\emph{Training.}---The residual $v_\theta$ is a convolutional network
with circular padding, time injected by a two-layer
SiLU MLP~\cite{Tan:2026ibs}. The SDE is discretized with $250$
uniform steps ($T=10$, $\Delta t=0.04$, $g=\sqrt2$) and trained with
Adam (learning rate $10^{-3}$, cosine-annealed to $10^{-5}$; batch
size 16) for $10^{3}$ epochs, with identical architecture and budget
for all $(L,\kappa)$.

\end{document}